\documentclass[%
 reprint,
 prl,
 amsmath,amssymb,
 aps,
 lengthcheck,
floatfix,
]{revtex4}

\usepackage{graphicx}% Include figure files
\usepackage{dcolumn}% Align table columns on decimal point
\usepackage{bm}% bold math
\usepackage{hyperref}% add hypertext capabilities
\usepackage[utf8]{inputenc}
\usepackage{soul}

\usepackage{graphicx}
\usepackage{times}
\usepackage[normalem]{ulem}
\usepackage{color}
\usepackage{cellspace}
\newcommand{\be}{\begin{equation}}
\newcommand{\ee}{\end{equation}}
\newcommand{\bea}{\begin{eqnarray}}
\newcommand{\eea}{\end{eqnarray}}

\newcommand{\comment}[1]{}
\renewcommand\sout{\bgroup \color{red} \ULdepth=-.5ex \ULset}
\def\simge{\mathrel{\rlap{\raise 0.511ex
     \hbox{$>$}}{\lower 0.511ex \hbox{$\sim$}}}}
\def\simle{\mathrel{\rlap{\raise 0.511ex
      \hbox{$<$}}{\lower 0.511ex \hbox{$\sim$}}}}

\begin{document}

% \setcounter{page}{1}
% \vspace*{0.3 true in}

\title{Tides in merging neutron stars: 
consistency of the GW170817 event with experimental data 
on finite nuclei}

% 
% and gravitational wave data} 

\author{Tuhin \surname{Malik} $^1$}
\email{tuhin.malik@gmail.com}
\author{B. K. \surname{Agrawal} $^{2,3}$}
%\email{bijay.agrawal@saha.ac.in}
\author{J.N. \surname{De}$^2$}
%\email{jn.de@saha.ac.in}
\author{S.K. \surname{Samaddar}$^2$}
%\email{santosh.samaddar@saha.ac.in}
\author{C. Provid{\^e}ncia$^4$}
%\email{cp@fis.uc.pt}
\author{C. Mondal$^5$}
\author{T.K. \surname{Jha}$^1$}
%\email{tkjha@goa.bits-pilani.ac.in}

\affiliation{$^1$Department of Physics, BITS-Pilani, K. K. Birla Goa Campus,
Goa 403726, India}
\affiliation{$^2$Saha Institute of Nuclear Physics, 1/AF 
 Bidhannagar, Kolkata 700064, India.}  
\affiliation{$^3$Homi Bhabha National Institute, Anushakti Nagar, Mumbai 400094, India.}  
\affiliation{$^4$CFisUC, Department of Physics, University of Coimbra, 3004-516 Coimbra, Portugal}
\affiliation{$^5$Departament de Física Quàntica i Astrofísica and Institut de Ciències del Cosmos (ICCUB), 
Facultat de Física, Universitat de Barcelona, Martí i Franquès 1, E-08028 Barcelona, Spain}

\date{\today}

\begin{abstract} 
The {agreement} of the nuclear equation of state (EoS) deduced from
the GW170817 based tidal deformability with the one  obtained from
empirical data on microscopic nuclei is examined. It is found that
suitably chosen experimental data on isoscalar and isovector modes of nuclear excitations together with the observed maximum neutron star
mass constrain the EoS which displays a very good congruence with the
GW170817 inspired one. The giant resonances in nuclei are found to
be  instrumental in limiting the tidal deformability parameter and the
radius of neutron star in somewhat narrower bounds. At the 1$\sigma$
level, the  values  of the canonical tidal deformability $\Lambda_{1.4}
$ and the neutron star radius $R_{1.4}$ come out to be  $267\pm144$
and $11.6\pm1.0$ km, respectively.

\end{abstract}

%\pacs{21.30.Fe, 21.65.Cd, 21.65.Mn, 21.65.Ef}
%\keywords{effective interaction}  

\maketitle
\textit{Introduction.}--- After the detection of gravitational waves
from the GW170817 binary neutron star merger event \cite{Abbott17},
the rich connection between the very large and the very small nuclear
objects has developed more intensely. During the last stages of the
inspiral motion of the coalescing neutron stars (NSs), the
{strong} gravity
of each of them induces a tidal deformation in the companion star. Decoding
the gravitational wave phase evolution caused by that deformation
\cite{Flanagan08} allows the determination of the  dimensionless tidal
deformability parameter $\Lambda $ \cite{Hinderer08, Hinderer10,Damour12}.
It is a measure of the  response to the gravitational pull on the neutron
star surface correlating with pressure gradients inside the NS and, therefore,
it has been proposed as an effective probe of the equation of state (EoS)
of nuclear matter relevant for neutron stars \cite{Thorne87,Read09}. A
comparatively large value of $\Lambda$, for example, points to a  neutron
star  of large radius \cite{Annala18,De18,Malik18}.  This translates
into a stiffer nuclear matter EoS and, hence, a comparatively larger neutron
skin of a heavy nucleus on the terrestrial plane \cite{Fattoyev18}.
Early analysis of the GW170817 event \cite{Abbott17} puts an upper
limit to the binary tidal deformability $\tilde \Lambda $ at $\approx
800$ 
for the component neutron stars with masses in the range $\approx
1.17-1.6 ~M_{\odot}$  involved in the merger event under the low spin
prior scenario.  $\tilde \Lambda $ is defined as
\bea
\label{lam1}
\tilde \Lambda=\frac{16}{13}\frac{(12q+1)\Lambda_1 +(12+q)q^4\Lambda_2}
{(1+q)^5},
\eea
where $\Lambda_{1,2}$ are the tidal deformabilities of the neutron
stars of masses $M_1$ and $M_2$ and $q=M_2/M_1 \leq 1$ is the binary's
mass ratio. The masses of the binary components are constrained
by the chirp mass ${\cal M}=(M_1M_2)^{3/5}/(M_1+M_2)^{1/5} = 1.188
M_{\odot}$ for GW170817 event, where $M_{\odot}$ is the solar mass. When $q=1$, $\tilde
\Lambda $ reduces to $\Lambda$ and is calculated from $\Lambda =
\frac{2}{3}k_2[\frac{c^2R}{GM}]^5$, where $k_2$ is the second Love number
\cite{Abbott17}, $R$ being the radius of the neutron star.  After the
initial proposition, the value of $\tilde \Lambda$ has gone through
several revisions  \cite{Abbott18a,Abbott18b,De18}.  Ref. \cite{De18}
reported $\tilde \Lambda =222^{+420}_{-138}$ for a uniform component-mass
prior at the 90\% credible level; with a few  plausible assumptions,
a  restrictive constraint is now set for a canonical  $\Lambda $
(=$\Lambda_{1.4}$, for a neutron star of mass $1.4 M_{\odot}$) at
$190^{+390}_{-120}$ \cite{Abbott18b} and the radii of both the lighter
and the heavier neutron stars in the merger event at $R_{1,2}=11.9\pm
1.4$ km.  From the spectral parameterization of the defining function
$p(\rho )$ ($p$=pressure) to fit the observational template, the
pressure inside the NS at supranormal densities is also predicted.
Complementing the electromagnetic probes that determine the maximum mass
of neutron stars ($2.01^{+0.04}_{-0.04} \le M_{\rm NS}^{\rm max}/M_\odot
\le 2.16^{+0.17}_{-0.15}$) \cite{Demorest10,Antoniadis13,Rezzolla18},
GW-based probes of the neutron star structure thus set the stage for exploring
the nuclear matter EoS at large densities.

First-principle calculations of nuclear matter EoS at subsaturation 
densities in chiral effective field theory (CEFT) \cite{Tews13} and at very 
high densities in perturbative QCD \cite{Kurkela10,Fraga16} are robust.
The problem of generating the most generic family of NS-matter EoS at
intermediate densities that interpolates between these reliable 
theoretical estimates consistent with the observational constraints  on
$M_{\rm NS}^{\rm max}$ and the tidal deformability has been recently
addressed \cite{Annala18}. A significant constraint on the
nuclear matter EoS is found from the inspection that the low density EoS must be stiff 
enough to support a NS of mass $\approx 2M_\odot$ but soft enough so 
that $\tilde \Lambda < 800$ \cite{Abbott18a}. Revisiting 
this problem with a huge number of parametrically constructed
plausible different EoSs connecting the low density and the high 
density end, Most $\it {et.al }$ \cite{Most18} find that,
for a purely hadronic star, the tidal deformability is constrained at 
$\Lambda_{1.4}>375$ at $2\sigma $ confidence level.
A non-parametric method for inferring the universal
neutron star matter EoS from GW observations is also reported 
recently \cite{Landry18} with  the canonical deformability
$\Lambda_{1.4}=160^{+448}_{-133}$ at 90$\% $ confidence level.
A lower bound on the tidal deformability $\approx 400$ is also set from the
analysis of the  UV-optical-infrared counterpart of GW170817 complemented with
numerical relativity results \cite{Radice17}. Similar
analysis, but, with a larger number of models pushes  the lower bound
to $\approx $ 200 \cite{Bauswein18}.

Through a combination of laboratory data on light nuclei and
sophisticated microscopic modeling of the sub-saturation EoS from CEFT
\cite{Tews18,Tews19,Lim18,Lim19}, attempts have been made to arrive at
values of the tidal deformability. Using a relativistic mean field (RMF)
inspired family of EoS models calibrated to provide a good description
of a set of 
selective properties of finite nuclei, the impact of the tidal deformability
on the  neutron-skin of $^{208}$Pb and on the NS mass and radius has also
been addressed \cite{Fattoyev18}. The varying outcomes point to the
fact that the connection of the tidal deformability to the laboratory
data is not yet fully transparent and that more stringent constraints
on the isovector sector of the effective interaction are needed. From
new-found strong correlations of $\Lambda_{1.4}$ and $R_{1.4}$ with a
set of
selective linear combinations of isoscalar and isovector properties of
nuclear matter, it is realized that such constraints may be provided
by the
isovector giant resonances in conjunction with the isoscalar resonances
in finite nuclei. To have a better understanding of these particularities,
in this communication, we perform an analysis of the suitability
of some often-used Skyrme models  to explain isoscalar
and isovector giant resonance data and examine their predictions for
$\Lambda_{1.4}$.
 {Simultaneously,  attention} is given to the
analysis of the astrophysical constraint on the neutron star maximum
mass $M_{NS}^{\rm max}$ \cite{Demorest10,Antoniadis13}; this encodes
pressure gradient information from mapping the varying neutron-proton
asymmetry over a large density range. Later, by fitting a broader-based
set of isoscalar and isovector data along with the observed NS mass
constraint, we propose a new EoS with the uncertainties estimated within
the covariance analysis and check its compatibility with the GW data.
The calculation is model dependent in the sense that the EoS is taken
to be a smooth function of density and avoids possibilities of phase
transitions to exotic form of matter when more drastic changes in the
density behavior of the EoS are considered.

\textit{Motivation from existing trends.}--- We resort to the Skyrme framework
for this study. For the suitability analysis of the Skyrme EDFs, we
choose among them twenty eight EDFs that are more representative. They include
the set of thirteen 'best' EDFs (CSkP set) used in Ref. \cite{Brown13}.
These are: KDE0v1, LNS, NRAPR, Ska25s20, Ska35s20, SKRA, SkT1, SkT2,
SkT3, SQMC700, Sv-sym32, Sly4, SkM*.  Another set of  thirteen Skyme
EDFs used in Ref.  \cite{Alam16} are also taken to examine the correlation
of the
neutron star radius with some key parameters of symmetric and asymmetric
nuclear matter. They are: Ska, Skb, SkI2, SkI3,
 SkI5, SkI6, Sly2, Sly230a, Sly9, SkMP, SkOP, SK255 and SK272.
To this list of twenty six, two recent EDFs, Sk$\chi $m$^*$ \cite{Zhang18}
and KDE0-J34 \cite{RocaMaza15} are further included; they are compliant
with the  measured dipole polarizability of few nuclei. The Sk$\chi
$m$^*$ EDF, in addition, reproduces  the theoretical predictions on
properties of asymmetric nuclear matter from  CEFT
\cite{Wellenhofer15,Wellenhofer16}. All these EDFs provide a
satisfactory reproduction of the binding energies of finite nuclei and
their charge radii, and obey reasonable constraints on the 
properties of symmetric nuclear matter such as the energy per nucleon ($e_0 =-15.8\pm 0.5$ MeV), the saturation
density ($\rho_0 =0.16\pm 0.01 ~\rm{fm}^{-3})$, the isoscalar nucleon
effective mass ($\frac{m_0^*}{m}=$0.6-1.0) and the isoscalar nuclear
incompressibility ( $K_0=240\pm 30$ MeV).

The twenty eight EDFs  mentioned above were constructed with emphasis on
different biases for the selection of data on finite nuclei and nuclear
matter properties.  We would like to have a closer look into these EDFs by
analyzing their ability to explain few further significant data related to
isoscalar and isovector properties of finite nuclei and draw inference
on the consistency of the EDFs in explaining observables concerning
neutron star masses and their tidal deformability.  The experimental
data of particular interest for  finite nuclei  are  the centroid energy
$E^{\rm{c}}_{\rm GMR}$ of the isoscalar giant monopole resonance (ISGMR),
the peak energy $E^{\rm{p}}_{\rm GDR}$ of the isovector giant dipole
resonance (IVGDR) and the dipole polarizability $\alpha_D$, all for the
heavy nucleus $^{208}$Pb.  The dipole polarizability $\alpha_D$ and the
GDR peak energies are measures of the isovector parameter $\Theta_v$ that
defines the isovector effective nucleon mass $m^*_{v,0}$ \cite{Zhang16}
in the Skyrme methodology. In conjunction with the isoscalar effective
mass $m_0^*$, this determines the isovector-splitting of the nucleon
effective mass $[\Delta m_0^* \equiv (m_n^*-m_p^*)/m]$, which is directly
related with 
the isovector properties of the nuclear interaction. Concerning the astrophysical context,
the data include the observed  lower limit of the maximum mass $M_{\rm
NS}^{\rm max}$ of the neutron star \cite{Demorest10,Antoniadis13},
($M_{\rm NS}^{\rm max}=2.01\pm 0.04 M_{\odot}$ ).

\begin{figure}
\includegraphics[width=1.0\columnwidth,angle=0,clip=true]{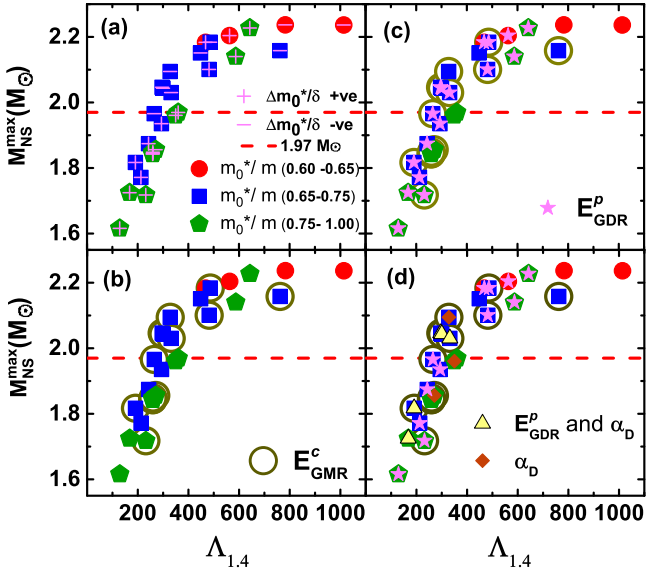}
\caption{\label{fig1}(color online) The maximum neutron star mass $M_{\rm
NS}^{\rm max}$ versus the tidal deformability parameter $\Lambda_{1.4}$
obtained from the 28 selected EDFs.  The red dashed lines refer to
$1.97M_\odot$, the observed lower bound for $M_{\rm NS}^{\rm max}$
. For more details , see text.} \end{figure}

The constraints provided by these empirical data allow to choose
the most plausible EDFs  considering the 
 neutron
star maximum mass and its radius, and the tidal deformability parameter
along with other properties of nuclear matter like $m_0^*$ or $\Delta
m_0^*$.  For the selected twenty eight EDFs, we find the effective mass
$\frac{m_0^*}{m}$ lying between $\approx $0.6-1.0  with $\Delta m_0^*$
distributed nearly evenly with positive and negative signs.  This is
shown as $(+)$ and $(-)$ signs for $\Delta m_0^*$ superimposed on the
symbols in  Fig.\ref{fig1}(a) where the calculated values of the maximum
neutron star mass $M_{\rm NS}^{\rm max}$ are given as a function of
the tidal deformability parameter $\Lambda_{1.4}$ for the given EDFs.
To focus on the role of $m_0^*$ in determining the ISGMR energy and
the maximum  mass of the neutron star, $\frac{m_0^*}{m}$ of the EDFs
are sorted in three groups, $0.60 \leq \frac{m_0^*}{m} < 0.65 $ (red
solid circle), $0.65 \leq \frac{m_0^*}{m} < 0.75$ (blue solid square)
and $0.75 \leq \frac{m_0^*}{m} \leq 1.0 $ (green solid pentagon).  The red
dashed horizontal lines in all the four panels in Fig.\ref{fig1} show
the lower bound of the  observed maximum value of the NS mass $ (=1.97
M_{\odot})$ that an acceptable EDF must support. To calculate the neutron
star properties, the EoS for its crust is taken from the Baym, Pethick and
Sutherland model \cite{Baym71} in the density range $\rho\approx 4.8\times
10^{-9}~\text{fm}^{-3}~\text{to}~2.6 \times 10^{-4}~\text{fm}^{-3}$. The
structure of the core is calculated from the EDFs with the assumption
of a charge neutral uniform plasma of neutrons, protons, electrons
and muons in $\beta$- equilibrium.  The EoS for the region between the
inner edge of the  outer crust and the  beginning of outer core defined by
the crust-core transition density is  appropriately interpolated using
a polytropic  form  \cite{Carriere03}.  This method may introduce
uncertainties in the determination of the radius of low and intermediate
NS masses \cite{Piekarewicz14,Fortin16,Pais16}. We have estimated an
average uncertainty of $\approx $2\% on $\Lambda_{1.4}$  by comparing the present
results with the ones obtained from unified EoSs. Fig.\ref{fig1}(a) shows
that the constraint on the NS maximum mass alone filters out some EDFs.
A good fraction of the  EDFs with effective masses above 0.75 $m$  fails
to achieve the lower bound on $M_{\rm NS}^{\rm max}$.

EDFs that fulfill the constraint imposed by the ISGMR centroid
energy in $^{208}$Pb (14.17$\pm $0.28 MeV ) are  represented by additional
open circle in Fig.\ref{fig1}(b). The  EDFs with effective masses in the
lower end of the spectrum (red solid circles, $\frac{m_0^*}{m} < 0.65 $)
are seen to be excluded from consideration; lower effective masses tend
to yield higher values of ISGMR energies than desired.  The further
constraint of satisfying the IVGDR peak energy (13.43 MeV; in
Ref. \cite{Dietrich88}, a large width of 4.07 MeV is ascribed to it. We
take a conservative estimate of 2 MeV for the width) for $^{208}$Pb
(marked with further  magenta-colored star) eliminates few more EDFs
as shown in Fig.\ref{fig1}(c) and, as is also seen there that, it forces the focus
on effective mass values in the middle range (0.65-0.75)$m$.
On top of these, imposition of the next constraint concerning the dipole
polarizability $\alpha_D$ for $^{208}$Pb (19.6 $\pm$0.6 fm$^3$) leaves
open the question of the suitability of most of the EDFs, as is seen
from the
inspection of Fig.\ref{fig1}(d). EDFs satisfying the constraint on
$\alpha_D$ are marked by orange diamonds, those satisfying criteria
concerning both the IVGDR peak energy and $\alpha_D$ are marked by yellow
triangles (see Table I of the Supplemental Material \cite{Suppl_prc} for details on 28 EDFs). Fig.\ref{fig1}(d) shows that among the selected twenty
eight EDFs, only three satisfy all the constraints
considered.  They are the  interactions Sly2, Sly4 and KDE0-J34. For these
three EDFs, the effective mass is $\approx 0.7 m$, and the isovector mass
splitting $\Delta m_0^*$ is negative.  It is of interest to note that the
constraints on the maximum NS mass and the ISGMR datum  in $^{208}$Pb can not
delineate the sign of the values of $\Delta m_0^*$, positive or negative;
the extra constraint on the peak energy of IVGDR in $^{208}$Pb is in favour of a negative $\Delta m_0^*$, the final constraint on
the dipole polarizability  settles this issue.  The value of the nucleon
effective mass (0.7$m$) is in very good agreement with that obtained
from the
optical model analysis of nucleon-nucleus scattering \cite{Li15}, but
the negative value of the isospin-splitted effective mass, at variance
with most theoretical predictions \cite{Li18,Li15,Zhang16,Kong17,Li13,
Holt16,Baldo17,Mondal17,Agrawal17}, needs possibly a more critical
examination.  Presently we do not discuss this matter except mentioning
that a recent EDF \cite{Malik18b} based on the Gibbs-Duhem relation and
specifically designed to fit a wide variety of 'pseudo data' corresponding
to infinite nuclear matter and the experimental energy weighted sum rule
for a few nuclei yields a value for the nucleon effective mass that is
very close ($\frac{m_0^*}{m} =0.68$) to what we find from this analysis
and also gives a negative value for $\Delta m_0^*$(=$-0.2 \delta $).
Here, $\delta$ is the isospin asymmetry of nuclear matter defined as $
\delta = (\rho_n-\rho_p)/\rho$, $\rho_n$ and $\rho_p$ being the neutron
and proton densities, respectively.

The role of the empirical data in sensitively constraining the tidal
deformability parameter $\Lambda$ should now be stressed.  One sees
from Fig.\ref{fig1} that from the total twenty eight EDFs chosen,
$\Lambda_{1.4}$ stretches out from 100 to 1000, the NS mass constraint
shrinks the band width to $\approx $ 270-1000, the ISGMR datum shrinks it
to $\approx $270-760, the IVGDR peak energy squeezes it further to $\approx
$270-590 and  $\Lambda_{1.4}$ settles it at $\approx $ 290-330 when filtered
through the choices of all the  data considered; it lies in midway of the
observed band width for $\Lambda_{1.4} $ deduced from the GW170817 event
\cite{Abbott18b}.  This survey suggests that there are models that can
endure the constraint on the observed $M_{\rm NS}^{\rm max}$, but many of
them would not fit the experimental data on the properties of the
ISGMR and IVGDR simultaneously due to the weak  correlations
among them as discussed later.  We would like to emphasize that the
conclusion drawn from Fig.\ref{fig1} is only indicative of the value
of the tidal deformability and serves as the motivation for the quantitative
investigation that follows.

\textit{Constraining tidal deformability from measured properties of finite nuclei.}---
To reassess the bounds on the tidal deformability more accurately, a new
Skyrme {EDF calibrated with a wider fit data base is proposed.}  
The constraints include the observed maximum NS mass $M_{\rm NS}^{\rm
max}$, the binding energies of spherical magic nuclei, their charge radii,
the ISGMR energy of $^{208}$Pb and its dipole polarizability. In addition,
the ISGMR energies of $^{90}$Zr and $^{120}$Sn and the dipole polarizibility  $\alpha_{\scriptstyle
D}$ of $^{48}$Ca, $^{68}$Ni and $^{120}$Sn are included in the fitting
protocol.

\begin{figure}
\includegraphics[width=1.0\columnwidth,angle=0,clip=true]{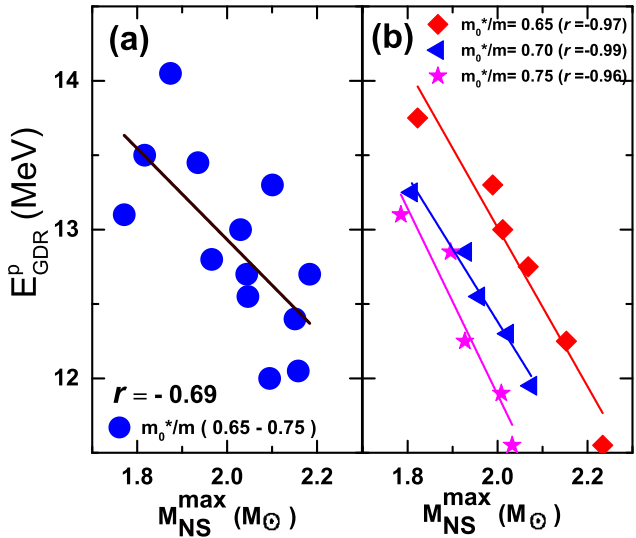}
\caption{\label{fig2}(color online) Correlation of $E_{\rm GDR}^{\rm p}$
and $M_{\rm NS}^{\rm max}$ obtained using (a) the set of selected models
as in Fig.\ref{fig1} with effective mass  $m^*_0/m$ in the range 0.65
-0.75 and (b) a set of systematically varied models with chosen fixed effective masses
in the present work.}
\end{figure}

\begin{table*}[t]
\caption{Parameters for the  model  Sk$\Lambda$267 and the resulting
nuclear matter and neutron star properties along with  their errors
in the parenthesis. $J_0$ is the symmetry energy coefficient, $L_0$
is related to its density derivative \cite{Malik18b}. } \label{tab1}
\setlength{\tabcolsep}{2.5pt}
\centering\begin{tabular}{cccccccccc} 
\hline
\rule{0pt}{3ex}%
$ t_0$ ( MeVfm$^3$ ) &$ t_1$ ( MeVfm$^5$ ) & $t_2$ ( MeVfm$^5$ ) &$ t_3
$( MeVfm$^{3+3 \alpha}$ ) & $x_0$ &$ x_1$ &$ x_2$ &$ x_3 $& $\alpha$ &$
W_0$ ( MeVfm$^5$ ) \\
\hline
%\rule{0pt}{3ex}%
 -2481.08 & 482.51 & -516.17 & 13778.74 & 0.93 & -0.53 & -0.97 & 1.54 & 0.167 & 121.38  \\
 (89.05) & (50.41) & (407.22) & (123.72) & (0.28) & (0.89) & (0.20) &  (0.58) & (0.018) &  (9.35) \\
\hline
 $e_0$(MeV) & $\rho_0$(fm$^{-3}) $& $K_0$(MeV) & $m_0^*/m $ &$J_0$(MeV)
&$L_0$(MeV)&$\Delta m_0^*/\delta$& $\Lambda_{1.4}$ & $R_{1.4}$ (km) & $M_{\rm NS}^{\rm
max}$ ($M_\odot$) \\
\hline
 $16.04$& 0.162& 230.2& 0.70&31.4& 41.1& -0.25&267 & 11.6& 2.04\\
( 0.20 )&  (0.002) &(6.4) & (0.05)&( 3.1) &( 18.2)& (0.35)& (144)&( 1.0) &(0.15)\\ 
\hline
\end{tabular}
\end{table*}
It is observed that  for the models employed in Fig.\ref{fig1},
$E_{\rm GMR}^{\rm c}$, $\alpha_{\scriptstyle D}$  and $M_{\rm NS}^{\rm
max}$ are weakly correlated among themselves (Pearson correlation
coefficients $ r$ are $ \approx 0.5$).  Simultaneously constraining
these quantities may impose strong restrictions on the model parameters.
The IVGDR peak energies are left out of the fitting protocol deliberately.
Calculations with the selected  EDFs reveal the existence of an
anti-correlation of $E_{\rm GDR}^{\rm p}$  for $^{208}$Pb with $M_{\rm
NS}^{\rm max}$ when the EDFs are sorted in groups within narrow windows
in $m_0^*/m$.  For illustration, this anti-correlation is displayed
in the Fig.\ref{fig2}(a) for  effective masses in the range $0.65 \le
m_0^*/m <0.75$ with the selected EDFs.  The correlation coefficient
is $r=-0.69$. However, we see that the aforesaid correlation shoots
up to nearly unity when calculated with the systematically varied
models obtained with fixed values of   $m_0^*/m$ as displayed in the
Fig. \ref{fig2}(b).  For given values of  $M_{\rm NS}^{\rm max}$ and
$m_0^*/m$, $E_{\rm GDR}^{\rm p}$ is the outcome of the calculation
keeping all other data in the fitting protocol unchanged.

The optimized $\chi^2$-function from the fit to all the input data
($M_{\rm NS}^{\rm max}$ and measured properties of finite nuclei as mentioned) yields
the EDF parameters. They are listed in Table \ref{tab1} along with
their errors obtained within the covariance method \cite{Dobaczewski14,
Mondal15, Zhang18}.  Some selected
properties of nuclear matter  and neutron stars are also presented in
the table. Since the central value of $\Lambda_{1.4}$ comes out to be
267, we hereafter {name this EDF} as Sk$\Lambda267$.  The nuclear matter
constants obtained for Sk$\Lambda267$ are in excellent agreement with
their fiducial values.  The lower bound on  $M_{\rm NS}^{\rm max}$ is
comfortably obeyed; the tidal deformability parameter ($\Lambda_{1.4}$)
and the NS radius $R_{1.4}$ are  also found to be in very good agreement
with that reported in Ref. \cite{Abbott18b}, the errors are more contained
though. The value of the neutron-skin $\Delta r_{np}$ for $^{208}$Pb is
$0.15\pm0.05$ fm. 

Since the experimental value of tidal deformability is
not yet settled, tolerance of the fit of the calculated observables with
the data is further tested by arbitrarily constraining  $\Lambda_{1.4}$ to
different values. As a demonstrative example we use an extra constraint in our fit
$\Lambda_{1.4}=500\pm100$. The outcome is model Sk$\Lambda$484 with
$\Lambda_{1.4}=484$ (see Table IV of the Supplemental Material \cite{Suppl_prc} for the
parameters). The model Sk$\Lambda$267 is found to be more compatible
with the measured properties of finite nuclei. A comparison of different observables  related
to nuclear matter and NS properties calculated with Sk$\Lambda$267 and
Sk$\Lambda$484 is given in Table III and IV of the Supplemental Material \cite{Suppl_prc}.
One  may note the closeness of the nuclear matter observables obtained
from Sk$\Lambda$267 and those from the interaction SLy4 \cite{Chabanat97}.
In SLy4, instead of the IVGDR as fit data as used in this paper, the
isotopic properties of forces beyond the $\beta$- stability line were
dictated by having a good reproduction of neutron-matter EoS obtained
variationally by Wiringa {\it et. al} \cite{Wiringa88,Wiringa93}.

\begin{figure}
\includegraphics[width=1.0\columnwidth,angle=0,clip=true]{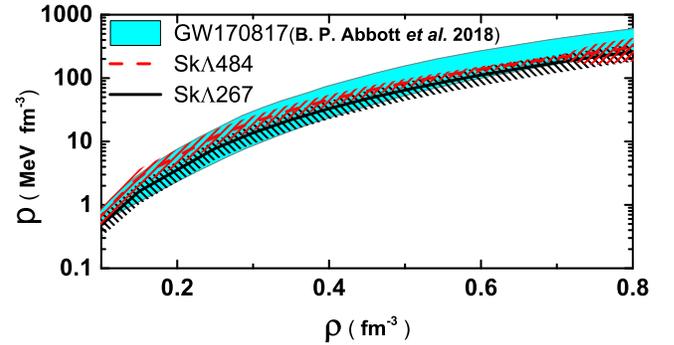}
\caption{\label{fig3}(color online) Pressure of $\beta$-equilibrated
neutron star matter displayed as a function of density. The shaded 
region represents the constraints from GW170817 event 
(B.P. Abbott {\it et. al} 2018: \cite{Abbott18b}).} 
\end{figure}
%The shaded
%region (cyan) shows results deduced from GW170817 event \cite{Abbott17}, the
%hatched region corresponds to the results obtained in the present work.}

The prediction of the EoSs Sk$\Lambda$267  and Sk$\Lambda$484 for
the pressure of the neutron star matter as a function of density  is
displayed in Fig.\ref{fig3} and compared with that deduced from the
GW170817 event \cite{Abbott18b}. As expected, Sk$\Lambda$267 is somewhat
softer than Sk$\Lambda$484.  Overall, the agreement between theory and
experiment is very good;  the delineation among the two theoretical EoSs
is, however, done through the microscopic lens  of the measured properties of finite nuclei
as already stated.  Both  EDFs maintain causality in the density range
encountered in the interior of the neutron stars; they become acausal
beyond $\rho\approx 8 \rho_0$ which is slightly higher than the central
density $\approx 7.0\rho_0$ for the maximum mass.

\textit{Remarks.}--- GW-based measurements of the macroscopic properties
of neutron stars offer a very promising means of looking deeper into
the nuclear microphysics governing the internal structure of the neutron
stars and of obtaining sound informative constraints on the nuclear
EoS at subnormal and supranormal densities.  We have explored in this
communication how the low density laboratory-data inspired nuclear matter
EoS connects with that obtained from GW-based data.  We show that
the pressure-density variation deduced from GW analysis is in very
good agreement with a parametric form of the EoS  designed to comply
with properly chosen nuclear observables sensitive to the
isoscalar and isovector parts of the nuclear interaction together
with the NS mass constraint.  The tidal deformability parameter is now
constrained at $\approx 267\pm144$ (267$\pm$ 236) at 1$\sigma $ level (90\%
confidence level).  We note that a recent reanalysis \cite{Narikawa18}
of the GW-based data leads to a considerable stretching of the bounds on
the tidal deformability although the central value ($\approx 200$)
maintains an
extremely good consistency with those  obtained earlier or with that
obtained by us. On  the other hand, the EoS derived from a neural network
\cite{Fujimoto19} having as input  observational data from several neutron
stars leads to $\Lambda_{1.4}= 320 \pm 120$ which is entirely consistent
with the values derived here.  Constraining NS properties from low-energy
nuclear physics thus seems very meaningful.  All nuclear properties, both
isoscalar and isovector, derived from our  EoS are in very comfortable
agreement with their fiducial values.  The values of the incompressibility,
the symmetry energy and its density derivative indicate that the EoS is
soft at densities near saturation; the conformity of the low value of the
tidal deformability with the most recent estimates shows that the EoS is
soft over a wider range of densities and  thus leaves the question open on
how to identify a possible phase transition in the neutron star core.
Future detections of binary star mergers by the LIGO-Virgo collaboration
may settle this issue.

\textit{Acknowledgments.}--- The authors acknowledge kind assistance
from Tanuja Agrawal in the preparation of the manuscript.
J.N.D. acknowledges support from the
Department of Science and Technology, Government of India  with
grant no. EMR/2016/001512. C.P. acknowledges financial support by
Fundação para a Ciência e Tecnologia (FCT) Portugal under project
No. UID/FIS/04564/2019, project POCI-01-0145-FEDER-029912 with financial
support from POCI, in its FEDER component, and by the FCT/MCTES budget
through national funds (OE), and the COST action CA16214 “PHAROS”.
C.M. acknowledges support from Project MDM-2014-0369 of ICCUB 
(Unidad de Exelencia María de Maeztu) from MINECO.

%\bibliography{prl}

\begin{thebibliography}{63}
\expandafter\ifx\csname natexlab\endcsname\relax\def\natexlab#1{#1}\fi
\expandafter\ifx\csname bibnamefont\endcsname\relax
  \def\bibnamefont#1{#1}\fi
\expandafter\ifx\csname bibfnamefont\endcsname\relax
  \def\bibfnamefont#1{#1}\fi
\expandafter\ifx\csname citenamefont\endcsname\relax
  \def\citenamefont#1{#1}\fi
\expandafter\ifx\csname url\endcsname\relax
  \def\url#1{\texttt{#1}}\fi
\expandafter\ifx\csname urlprefix\endcsname\relax\def\urlprefix{URL }\fi
\providecommand{\bibinfo}[2]{#2}
\providecommand{\eprint}[2][]{\url{#2}}

\bibitem[{\citenamefont{Abbott et~al.}(2017)\citenamefont{Abbott, Abbott,
  Abbott, Acernese, Ackley, Adams, Adams, Addesso, Adhikari, Adya
  et~al.}}]{Abbott17}
\bibinfo{author}{\bibfnamefont{B.~P.} \bibnamefont{Abbott}},
  \bibinfo{author}{\bibfnamefont{R.}~\bibnamefont{Abbott}},
  \bibinfo{author}{\bibfnamefont{T.~D.} \bibnamefont{Abbott}},
  \bibinfo{author}{\bibfnamefont{F.}~\bibnamefont{Acernese}},
  \bibinfo{author}{\bibfnamefont{K.}~\bibnamefont{Ackley}},
  \bibinfo{author}{\bibfnamefont{C.}~\bibnamefont{Adams}},
  \bibinfo{author}{\bibfnamefont{T.}~\bibnamefont{Adams}},
  \bibinfo{author}{\bibfnamefont{P.}~\bibnamefont{Addesso}},
  \bibinfo{author}{\bibfnamefont{R.~X.} \bibnamefont{Adhikari}},
  \bibinfo{author}{\bibfnamefont{V.~B.} \bibnamefont{Adya}},
  \bibnamefont{et~al.}, \bibinfo{journal}{Phys. Rev. Lett.}
  \textbf{\bibinfo{volume}{119}}, \bibinfo{pages}{161101}
  (\bibinfo{year}{2017}).

\bibitem[{\citenamefont{Flanagan and Hinderer}(2008)}]{Flanagan08}
\bibinfo{author}{\bibfnamefont{E.~E.} \bibnamefont{Flanagan}} \bibnamefont{and}
  \bibinfo{author}{\bibfnamefont{T.}~\bibnamefont{Hinderer}},
  \bibinfo{journal}{Phys. Rev. D} \textbf{\bibinfo{volume}{77}},
  \bibinfo{pages}{021502} (\bibinfo{year}{2008}).

\bibitem[{\citenamefont{Hinderer}(2008)}]{Hinderer08}
\bibinfo{author}{\bibfnamefont{T.}~\bibnamefont{Hinderer}},
  \bibinfo{journal}{Astrophys. J.} \textbf{\bibinfo{volume}{677}},
  \bibinfo{pages}{1216} (\bibinfo{year}{2008}).

\bibitem[{\citenamefont{Hinderer et~al.}(2010)\citenamefont{Hinderer, Lackey,
  Lang, and Read}}]{Hinderer10}
\bibinfo{author}{\bibfnamefont{T.}~\bibnamefont{Hinderer}},
  \bibinfo{author}{\bibfnamefont{B.~D.} \bibnamefont{Lackey}},
  \bibinfo{author}{\bibfnamefont{R.~N.} \bibnamefont{Lang}}, \bibnamefont{and}
  \bibinfo{author}{\bibfnamefont{J.~S.} \bibnamefont{Read}},
  \bibinfo{journal}{Phys. Rev. D} \textbf{\bibinfo{volume}{81}},
  \bibinfo{pages}{123016} (\bibinfo{year}{2010}).

\bibitem[{\citenamefont{Damour et~al.}(2012)\citenamefont{Damour, Nagar, and
  Villain}}]{Damour12}
\bibinfo{author}{\bibfnamefont{T.}~\bibnamefont{Damour}},
  \bibinfo{author}{\bibfnamefont{A.}~\bibnamefont{Nagar}}, \bibnamefont{and}
  \bibinfo{author}{\bibfnamefont{L.}~\bibnamefont{Villain}},
  \bibinfo{journal}{Phys. Rev. D} \textbf{\bibinfo{volume}{85}},
  \bibinfo{pages}{123007} (\bibinfo{year}{2012}).

\bibitem[{\citenamefont{Thorne}((1987))}]{Thorne87}
\bibinfo{author}{\bibfnamefont{K.}~\bibnamefont{Thorne}},
  \emph{\bibinfo{title}{{{\it Three Hundred Years of Gravitation}}}}
  (\bibinfo{publisher}{Cambridge, UK: Univ. Pr. 684 p},
  \bibinfo{year}{(1987)}).

\bibitem[{\citenamefont{Read et~al.}(2009)\citenamefont{Read, Markakis,
  Shibata, Ury\ifmmode~\bar{u}\else \={u}\fi{}, Creighton, and
  Friedman}}]{Read09}
\bibinfo{author}{\bibfnamefont{J.~S.} \bibnamefont{Read}},
  \bibinfo{author}{\bibfnamefont{C.}~\bibnamefont{Markakis}},
  \bibinfo{author}{\bibfnamefont{M.}~\bibnamefont{Shibata}},
  \bibinfo{author}{\bibfnamefont{K.~b.~o.}
  \bibnamefont{Ury\ifmmode~\bar{u}\else \={u}\fi{}}},
  \bibinfo{author}{\bibfnamefont{J.~D.~E.} \bibnamefont{Creighton}},
  \bibnamefont{and} \bibinfo{author}{\bibfnamefont{J.~L.}
  \bibnamefont{Friedman}}, \bibinfo{journal}{Phys. Rev. D}
  \textbf{\bibinfo{volume}{79}}, \bibinfo{pages}{124033}
  (\bibinfo{year}{2009}).

\bibitem[{\citenamefont{Annala et~al.}(2018)\citenamefont{Annala, Gorda,
  Kurkela, and Vuorinen}}]{Annala18}
\bibinfo{author}{\bibfnamefont{E.}~\bibnamefont{Annala}},
  \bibinfo{author}{\bibfnamefont{T.}~\bibnamefont{Gorda}},
  \bibinfo{author}{\bibfnamefont{A.}~\bibnamefont{Kurkela}}, \bibnamefont{and}
  \bibinfo{author}{\bibfnamefont{A.}~\bibnamefont{Vuorinen}},
  \bibinfo{journal}{Phys. Rev. Lett.} \textbf{\bibinfo{volume}{120}},
  \bibinfo{pages}{172703} (\bibinfo{year}{2018}).

\bibitem[{\citenamefont{De et~al.}(2018)\citenamefont{De, Finstad, Lattimer,
  Brown, Berger, and Biwer}}]{De18}
\bibinfo{author}{\bibfnamefont{S.}~\bibnamefont{De}},
  \bibinfo{author}{\bibfnamefont{D.}~\bibnamefont{Finstad}},
  \bibinfo{author}{\bibfnamefont{J.~M.} \bibnamefont{Lattimer}},
  \bibinfo{author}{\bibfnamefont{D.~A.} \bibnamefont{Brown}},
  \bibinfo{author}{\bibfnamefont{E.}~\bibnamefont{Berger}}, \bibnamefont{and}
  \bibinfo{author}{\bibfnamefont{C.~M.} \bibnamefont{Biwer}},
  \bibinfo{journal}{Phys. Rev. Lett.} \textbf{\bibinfo{volume}{121}},
  \bibinfo{pages}{091102} (\bibinfo{year}{2018}).

\bibitem[{\citenamefont{Malik et~al.}(2018{\natexlab{a}})\citenamefont{Malik,
  Alam, Fortin, Provid\^encia, Agrawal, Jha, Kumar, and Patra}}]{Malik18}
\bibinfo{author}{\bibfnamefont{T.}~\bibnamefont{Malik}},
  \bibinfo{author}{\bibfnamefont{N.}~\bibnamefont{Alam}},
  \bibinfo{author}{\bibfnamefont{M.}~\bibnamefont{Fortin}},
  \bibinfo{author}{\bibfnamefont{C.}~\bibnamefont{Provid\^encia}},
  \bibinfo{author}{\bibfnamefont{B.~K.} \bibnamefont{Agrawal}},
  \bibinfo{author}{\bibfnamefont{T.~K.} \bibnamefont{Jha}},
  \bibinfo{author}{\bibfnamefont{B.}~\bibnamefont{Kumar}}, \bibnamefont{and}
  \bibinfo{author}{\bibfnamefont{S.~K.} \bibnamefont{Patra}},
  \bibinfo{journal}{Phys. Rev. C} \textbf{\bibinfo{volume}{98}},
  \bibinfo{pages}{035804} (\bibinfo{year}{2018}{\natexlab{a}}).

\bibitem[{\citenamefont{Fattoyev et~al.}(2018)\citenamefont{Fattoyev,
  Piekarewicz, and Horowitz}}]{Fattoyev18}
\bibinfo{author}{\bibfnamefont{F.~J.} \bibnamefont{Fattoyev}},
  \bibinfo{author}{\bibfnamefont{J.}~\bibnamefont{Piekarewicz}},
  \bibnamefont{and} \bibinfo{author}{\bibfnamefont{C.~J.}
  \bibnamefont{Horowitz}}, \bibinfo{journal}{Phys. Rev. Lett.}
  \textbf{\bibinfo{volume}{120}}, \bibinfo{pages}{172702}
  (\bibinfo{year}{2018}).

\bibitem[{\citenamefont{Abbott et~al.}(2019)}]{Abbott18a}
\bibinfo{author}{\bibfnamefont{B.~P.} \bibnamefont{Abbott}}
  \bibnamefont{et~al.} (\bibinfo{collaboration}{LIGO Scientific, Virgo}),
  \bibinfo{journal}{Phys. Rev.} \textbf{\bibinfo{volume}{X9}},
  \bibinfo{pages}{011001} (\bibinfo{year}{2019}).

\bibitem[{\citenamefont{Abbott et~al.}(2018)\citenamefont{Abbott, Abbott,
  Abbott, Acernese, Ackley, Adams, Adams, Addesso, Adhikari, Adya
  et~al.}}]{Abbott18b}
\bibinfo{author}{\bibfnamefont{B.~P.} \bibnamefont{Abbott}},
  \bibinfo{author}{\bibfnamefont{R.}~\bibnamefont{Abbott}},
  \bibinfo{author}{\bibfnamefont{T.~D.} \bibnamefont{Abbott}},
  \bibinfo{author}{\bibfnamefont{F.}~\bibnamefont{Acernese}},
  \bibinfo{author}{\bibfnamefont{K.}~\bibnamefont{Ackley}},
  \bibinfo{author}{\bibfnamefont{C.}~\bibnamefont{Adams}},
  \bibinfo{author}{\bibfnamefont{T.}~\bibnamefont{Adams}},
  \bibinfo{author}{\bibfnamefont{P.}~\bibnamefont{Addesso}},
  \bibinfo{author}{\bibfnamefont{R.~X.} \bibnamefont{Adhikari}},
  \bibinfo{author}{\bibfnamefont{V.~B.} \bibnamefont{Adya}},
  \bibnamefont{et~al.}, \bibinfo{journal}{Phys. Rev. Lett.}
  \textbf{\bibinfo{volume}{121}}, \bibinfo{pages}{161101}
  (\bibinfo{year}{2018}).

\bibitem[{\citenamefont{Demorest et~al.}(2010)\citenamefont{Demorest, Pennucci,
  Ransom, Roberts, and Hessels}}]{Demorest10}
\bibinfo{author}{\bibfnamefont{P.}~\bibnamefont{Demorest}},
  \bibinfo{author}{\bibfnamefont{T.}~\bibnamefont{Pennucci}},
  \bibinfo{author}{\bibfnamefont{S.}~\bibnamefont{Ransom}},
  \bibinfo{author}{\bibfnamefont{M.}~\bibnamefont{Roberts}}, \bibnamefont{and}
  \bibinfo{author}{\bibfnamefont{J.}~\bibnamefont{Hessels}},
  \bibinfo{journal}{Nature} \textbf{\bibinfo{volume}{467}},
  \bibinfo{pages}{1081} (\bibinfo{year}{2010}), \eprint{1010.5788}.

\bibitem[{\citenamefont{{Antoniadis} et~al.}(2013)\citenamefont{{Antoniadis},
  {Freire}, {Wex}, {Tauris}, {Lynch}, {van Kerkwijk}, {Kramer}, {Bassa},
  {Dhillon}, {Driebe} et~al.}}]{Antoniadis13}
\bibinfo{author}{\bibfnamefont{J.}~\bibnamefont{{Antoniadis}}},
  \bibinfo{author}{\bibfnamefont{P.~C.~C.} \bibnamefont{{Freire}}},
  \bibinfo{author}{\bibfnamefont{N.}~\bibnamefont{{Wex}}},
  \bibinfo{author}{\bibfnamefont{T.~M.} \bibnamefont{{Tauris}}},
  \bibinfo{author}{\bibfnamefont{R.~S.} \bibnamefont{{Lynch}}},
  \bibinfo{author}{\bibfnamefont{M.~H.} \bibnamefont{{van Kerkwijk}}},
  \bibinfo{author}{\bibfnamefont{M.}~\bibnamefont{{Kramer}}},
  \bibinfo{author}{\bibfnamefont{C.}~\bibnamefont{{Bassa}}},
  \bibinfo{author}{\bibfnamefont{V.~S.} \bibnamefont{{Dhillon}}},
  \bibinfo{author}{\bibfnamefont{T.}~\bibnamefont{{Driebe}}},
  \bibnamefont{et~al.}, \bibinfo{journal}{Science}
  \textbf{\bibinfo{volume}{340}}, \bibinfo{pages}{448} (\bibinfo{year}{2013}),
  \eprint{1304.6875}.

\bibitem[{\citenamefont{Rezzolla et~al.}(2018)\citenamefont{Rezzolla, Most, and
  Weih}}]{Rezzolla18}
\bibinfo{author}{\bibfnamefont{L.}~\bibnamefont{Rezzolla}},
  \bibinfo{author}{\bibfnamefont{E.~R.} \bibnamefont{Most}}, \bibnamefont{and}
  \bibinfo{author}{\bibfnamefont{L.~R.} \bibnamefont{Weih}},
  \bibinfo{journal}{Astrophys. J. Lett.} \textbf{\bibinfo{volume}{852}},
  \bibinfo{pages}{L25} (\bibinfo{year}{2018}).

\bibitem[{\citenamefont{Tews et~al.}(2013)\citenamefont{Tews, Kr\"uger,
  Hebeler, and Schwenk}}]{Tews13}
\bibinfo{author}{\bibfnamefont{I.}~\bibnamefont{Tews}},
  \bibinfo{author}{\bibfnamefont{T.}~\bibnamefont{Kr\"uger}},
  \bibinfo{author}{\bibfnamefont{K.}~\bibnamefont{Hebeler}}, \bibnamefont{and}
  \bibinfo{author}{\bibfnamefont{A.}~\bibnamefont{Schwenk}},
  \bibinfo{journal}{Phys. Rev. Lett.} \textbf{\bibinfo{volume}{110}},
  \bibinfo{pages}{032504} (\bibinfo{year}{2013}).

\bibitem[{\citenamefont{Kurkela et~al.}(2010)\citenamefont{Kurkela, Romatschke,
  and Vuorinen}}]{Kurkela10}
\bibinfo{author}{\bibfnamefont{A.}~\bibnamefont{Kurkela}},
  \bibinfo{author}{\bibfnamefont{P.}~\bibnamefont{Romatschke}},
  \bibnamefont{and} \bibinfo{author}{\bibfnamefont{A.}~\bibnamefont{Vuorinen}},
  \bibinfo{journal}{Phys. Rev. D} \textbf{\bibinfo{volume}{81}},
  \bibinfo{pages}{105021} (\bibinfo{year}{2010}).

\bibitem[{\citenamefont{Fraga et~al.}(2016)\citenamefont{Fraga, Kurkela, and
  Vuorinen}}]{Fraga16}
\bibinfo{author}{\bibfnamefont{E.~S.} \bibnamefont{Fraga}},
  \bibinfo{author}{\bibfnamefont{A.}~\bibnamefont{Kurkela}}, \bibnamefont{and}
  \bibinfo{author}{\bibfnamefont{A.}~\bibnamefont{Vuorinen}},
  \bibinfo{journal}{Eur. Phys. J.} \textbf{\bibinfo{volume}{A52}},
  \bibinfo{pages}{49} (\bibinfo{year}{2016}), \eprint{1508.05019}.

\bibitem[{\citenamefont{Most et~al.}(2018)\citenamefont{Most, Weih, Rezzolla,
  and Schaffner-Bielich}}]{Most18}
\bibinfo{author}{\bibfnamefont{E.~R.} \bibnamefont{Most}},
  \bibinfo{author}{\bibfnamefont{L.~R.} \bibnamefont{Weih}},
  \bibinfo{author}{\bibfnamefont{L.}~\bibnamefont{Rezzolla}}, \bibnamefont{and}
  \bibinfo{author}{\bibfnamefont{J.}~\bibnamefont{Schaffner-Bielich}},
  \bibinfo{journal}{Phys. Rev. Lett.} \textbf{\bibinfo{volume}{120}},
  \bibinfo{pages}{261103} (\bibinfo{year}{2018}).

\bibitem[{\citenamefont{Landry and Essick}(2018)}]{Landry18}
\bibinfo{author}{\bibfnamefont{P.}~\bibnamefont{Landry}} \bibnamefont{and}
  \bibinfo{author}{\bibfnamefont{R.}~\bibnamefont{Essick}},
  \bibinfo{journal}{arXiv:1811.12529}  (\bibinfo{year}{2018}).

\bibitem[{\citenamefont{Radice et~al.}(2018)\citenamefont{Radice, Perego,
  Zappa, and Bernuzzi}}]{Radice17}
\bibinfo{author}{\bibfnamefont{D.}~\bibnamefont{Radice}},
  \bibinfo{author}{\bibfnamefont{A.}~\bibnamefont{Perego}},
  \bibinfo{author}{\bibfnamefont{F.}~\bibnamefont{Zappa}}, \bibnamefont{and}
  \bibinfo{author}{\bibfnamefont{S.}~\bibnamefont{Bernuzzi}},
  \bibinfo{journal}{Astrophys. J. Lett.} \textbf{\bibinfo{volume}{852}},
  \bibinfo{pages}{L29} (\bibinfo{year}{2018}).

\bibitem[{\citenamefont{Bauswein}(2018)}]{Bauswein18}
\bibinfo{author}{\bibfnamefont{A.}~\bibnamefont{Bauswein}},
  \bibinfo{journal}{Talk delivered at International School of Nuclear Physics}
  (\bibinfo{year}{2018}),
  \urlprefix\url{crunch.ikp.physik.tu-darmstadt.de/erice/2018/sec/talks/sunday/bauswein.pdf}.

\bibitem[{\citenamefont{Tews et~al.}(2018)\citenamefont{Tews, Margueron, and
  Reddy}}]{Tews18}
\bibinfo{author}{\bibfnamefont{I.}~\bibnamefont{Tews}},
  \bibinfo{author}{\bibfnamefont{J.}~\bibnamefont{Margueron}},
  \bibnamefont{and} \bibinfo{author}{\bibfnamefont{S.}~\bibnamefont{Reddy}},
  \bibinfo{journal}{Phys. Rev. C} \textbf{\bibinfo{volume}{98}},
  \bibinfo{pages}{045804} (\bibinfo{year}{2018}).

\bibitem[{\citenamefont{Tews et~al.}(2019)\citenamefont{Tews, Margueron, and
  Reddy}}]{Tews19}
\bibinfo{author}{\bibfnamefont{I.}~\bibnamefont{Tews}},
  \bibinfo{author}{\bibfnamefont{J.}~\bibnamefont{Margueron}},
  \bibnamefont{and} \bibinfo{author}{\bibfnamefont{S.}~\bibnamefont{Reddy}},
  \bibinfo{journal}{arXiv:1901.09874 [nucl-th]}  (\bibinfo{year}{2019}).

\bibitem[{\citenamefont{Lim and Holt}(2018)}]{Lim18}
\bibinfo{author}{\bibfnamefont{Y.}~\bibnamefont{Lim}} \bibnamefont{and}
  \bibinfo{author}{\bibfnamefont{J.~W.} \bibnamefont{Holt}},
  \bibinfo{journal}{Phys. Rev. Lett.} \textbf{\bibinfo{volume}{121}},
  \bibinfo{pages}{062701} (\bibinfo{year}{2018}).

\bibitem[{\citenamefont{Lim and Holt}(2019)}]{Lim19}
\bibinfo{author}{\bibfnamefont{Y.}~\bibnamefont{Lim}} \bibnamefont{and}
  \bibinfo{author}{\bibfnamefont{J.~W.} \bibnamefont{Holt}},
  \bibinfo{journal}{arXiv:1902.05502}  (\bibinfo{year}{2019}).

\bibitem[{\citenamefont{Brown}(2013)}]{Brown13}
\bibinfo{author}{\bibfnamefont{B.~A.} \bibnamefont{Brown}},
  \bibinfo{journal}{Phys. Rev. Lett.} \textbf{\bibinfo{volume}{111}},
  \bibinfo{pages}{232502} (\bibinfo{year}{2013}).

\bibitem[{\citenamefont{Alam et~al.}(2016)\citenamefont{Alam, Agrawal, Fortin,
  Pais, Provid\^encia, Raduta, and Sulaksono}}]{Alam16}
\bibinfo{author}{\bibfnamefont{N.}~\bibnamefont{Alam}},
  \bibinfo{author}{\bibfnamefont{B.~K.} \bibnamefont{Agrawal}},
  \bibinfo{author}{\bibfnamefont{M.}~\bibnamefont{Fortin}},
  \bibinfo{author}{\bibfnamefont{H.}~\bibnamefont{Pais}},
  \bibinfo{author}{\bibfnamefont{C.}~\bibnamefont{Provid\^encia}},
  \bibinfo{author}{\bibfnamefont{A.~R.} \bibnamefont{Raduta}},
  \bibnamefont{and}
  \bibinfo{author}{\bibfnamefont{A.}~\bibnamefont{Sulaksono}},
  \bibinfo{journal}{Phys. Rev. C} \textbf{\bibinfo{volume}{94}},
  \bibinfo{pages}{052801} (\bibinfo{year}{2016}).

\bibitem[{\citenamefont{Zhang et~al.}(2018)\citenamefont{Zhang, Lim, Holt, and
  Ko}}]{Zhang18}
\bibinfo{author}{\bibfnamefont{Z.}~\bibnamefont{Zhang}},
  \bibinfo{author}{\bibfnamefont{Y.}~\bibnamefont{Lim}},
  \bibinfo{author}{\bibfnamefont{J.~W.} \bibnamefont{Holt}}, \bibnamefont{and}
  \bibinfo{author}{\bibfnamefont{C.~M.} \bibnamefont{Ko}},
  \bibinfo{journal}{Physics Letters B} \textbf{\bibinfo{volume}{777}},
  \bibinfo{pages}{73 } (\bibinfo{year}{2018}).

\bibitem[{\citenamefont{Roca-Maza et~al.}(2015)\citenamefont{Roca-Maza,
  Vi\~nas, Centelles, Agrawal, Col\`o, Paar, Piekarewicz, and
  Vretenar}}]{RocaMaza15}
\bibinfo{author}{\bibfnamefont{X.}~\bibnamefont{Roca-Maza}},
  \bibinfo{author}{\bibfnamefont{X.}~\bibnamefont{Vi\~nas}},
  \bibinfo{author}{\bibfnamefont{M.}~\bibnamefont{Centelles}},
  \bibinfo{author}{\bibfnamefont{B.~K.} \bibnamefont{Agrawal}},
  \bibinfo{author}{\bibfnamefont{G.}~\bibnamefont{Col\`o}},
  \bibinfo{author}{\bibfnamefont{N.}~\bibnamefont{Paar}},
  \bibinfo{author}{\bibfnamefont{J.}~\bibnamefont{Piekarewicz}},
  \bibnamefont{and} \bibinfo{author}{\bibfnamefont{D.}~\bibnamefont{Vretenar}},
  \bibinfo{journal}{Phys. Rev. C} \textbf{\bibinfo{volume}{92}},
  \bibinfo{pages}{064304} (\bibinfo{year}{2015}).

\bibitem[{\citenamefont{Wellenhofer et~al.}(2015)\citenamefont{Wellenhofer,
  Holt, and Kaiser}}]{Wellenhofer15}
\bibinfo{author}{\bibfnamefont{C.}~\bibnamefont{Wellenhofer}},
  \bibinfo{author}{\bibfnamefont{J.~W.} \bibnamefont{Holt}}, \bibnamefont{and}
  \bibinfo{author}{\bibfnamefont{N.}~\bibnamefont{Kaiser}},
  \bibinfo{journal}{Phys. Rev. C} \textbf{\bibinfo{volume}{92}},
  \bibinfo{pages}{015801} (\bibinfo{year}{2015}).

\bibitem[{\citenamefont{Wellenhofer et~al.}(2016)\citenamefont{Wellenhofer,
  Holt, and Kaiser}}]{Wellenhofer16}
\bibinfo{author}{\bibfnamefont{C.}~\bibnamefont{Wellenhofer}},
  \bibinfo{author}{\bibfnamefont{J.~W.} \bibnamefont{Holt}}, \bibnamefont{and}
  \bibinfo{author}{\bibfnamefont{N.}~\bibnamefont{Kaiser}},
  \bibinfo{journal}{Phys. Rev. C} \textbf{\bibinfo{volume}{93}},
  \bibinfo{pages}{055802} (\bibinfo{year}{2016}).

\bibitem[{\citenamefont{Zhang and Chen}(2016)}]{Zhang16}
\bibinfo{author}{\bibfnamefont{Z.}~\bibnamefont{Zhang}} \bibnamefont{and}
  \bibinfo{author}{\bibfnamefont{L.-W.} \bibnamefont{Chen}},
  \bibinfo{journal}{Phys. Rev. C} \textbf{\bibinfo{volume}{93}},
  \bibinfo{pages}{034335} (\bibinfo{year}{2016}).

\bibitem[{\citenamefont{{Baym} et~al.}(1971)\citenamefont{{Baym}, {Pethick},
  and {Sutherland}}}]{Baym71}
\bibinfo{author}{\bibfnamefont{G.}~\bibnamefont{{Baym}}},
  \bibinfo{author}{\bibfnamefont{C.}~\bibnamefont{{Pethick}}},
  \bibnamefont{and}
  \bibinfo{author}{\bibfnamefont{P.}~\bibnamefont{{Sutherland}}},
  \bibinfo{journal}{\apj} \textbf{\bibinfo{volume}{170}}, \bibinfo{pages}{299}
  (\bibinfo{year}{1971}).

\bibitem[{\citenamefont{Carriere et~al.}(2003)\citenamefont{Carriere, Horowitz,
  and Piekarewicz}}]{Carriere03}
\bibinfo{author}{\bibfnamefont{J.}~\bibnamefont{Carriere}},
  \bibinfo{author}{\bibfnamefont{C.}~\bibnamefont{Horowitz}}, \bibnamefont{and}
  \bibinfo{author}{\bibfnamefont{J.}~\bibnamefont{Piekarewicz}},
  \bibinfo{journal}{Astrophys. J.} \textbf{\bibinfo{volume}{593}},
  \bibinfo{pages}{463} (\bibinfo{year}{2003}).

\bibitem[{\citenamefont{Piekarewicz et~al.}(2014)\citenamefont{Piekarewicz,
  Fattoyev, and Horowitz}}]{Piekarewicz14}
\bibinfo{author}{\bibfnamefont{J.}~\bibnamefont{Piekarewicz}},
  \bibinfo{author}{\bibfnamefont{F.~J.} \bibnamefont{Fattoyev}},
  \bibnamefont{and} \bibinfo{author}{\bibfnamefont{C.~J.}
  \bibnamefont{Horowitz}}, \bibinfo{journal}{Phys. Rev. C}
  \textbf{\bibinfo{volume}{90}}, \bibinfo{pages}{015803}
  (\bibinfo{year}{2014}).

\bibitem[{\citenamefont{Fortin et~al.}(2016)\citenamefont{Fortin, Providencia,
  Raduta, Gulminelli, Zdunik, Haensel, and Bejger}}]{Fortin16}
\bibinfo{author}{\bibfnamefont{M.}~\bibnamefont{Fortin}},
  \bibinfo{author}{\bibfnamefont{C.}~\bibnamefont{Providencia}},
  \bibinfo{author}{\bibfnamefont{A.~R.} \bibnamefont{Raduta}},
  \bibinfo{author}{\bibfnamefont{F.}~\bibnamefont{Gulminelli}},
  \bibinfo{author}{\bibfnamefont{J.~L.} \bibnamefont{Zdunik}},
  \bibinfo{author}{\bibfnamefont{P.}~\bibnamefont{Haensel}}, \bibnamefont{and}
  \bibinfo{author}{\bibfnamefont{M.}~\bibnamefont{Bejger}},
  \bibinfo{journal}{Phys. Rev.} \textbf{\bibinfo{volume}{C94}},
  \bibinfo{pages}{035804} (\bibinfo{year}{2016}), \eprint{1604.01944}.

\bibitem[{\citenamefont{Pais and Providência}(2016)}]{Pais16}
\bibinfo{author}{\bibfnamefont{H.}~\bibnamefont{Pais}} \bibnamefont{and}
  \bibinfo{author}{\bibfnamefont{C.}~\bibnamefont{Providência}},
  \bibinfo{journal}{Phys. Rev.} \textbf{\bibinfo{volume}{C94}},
  \bibinfo{pages}{015808} (\bibinfo{year}{2016}), \eprint{1607.05899}.

\bibitem[{\citenamefont{Dietrich and Berman}(1988)}]{Dietrich88}
\bibinfo{author}{\bibfnamefont{S.~S.} \bibnamefont{Dietrich}} \bibnamefont{and}
  \bibinfo{author}{\bibfnamefont{B.}~\bibnamefont{Berman}},
  \bibinfo{journal}{At. Data. Nucl. Data Tables} \textbf{\bibinfo{volume}{38}},
  \bibinfo{pages}{199} (\bibinfo{year}{1988}).

\bibitem[{Sup()}]{Suppl_prc}
\bibinfo{note}{See Supplemental Material at https://journals.aps.org/prc/supplemental/10.1103/PhysRevC.99.052801 for the compilation of the
  properties of finite nuclei, nuclear matter and neutron star corresponding to
  the 28 representatives EDFs as used in the present work along with the result
  obtained for the newly generated Skyrme forces Sk$\Lambda267$ and
  Sk$\Lambda484$, which includes Refs.
  \cite{Fuchs06,Fantina14,Danielewicz02,Hebeler13,Colo13,RocaMaza15,Dutra12}}.

\bibitem[{\citenamefont{Li et~al.}(2015)\citenamefont{Li, Guo, Li, Chen,
  Fattoyev, and Newton}}]{Li15}
\bibinfo{author}{\bibfnamefont{X.-H.} \bibnamefont{Li}},
  \bibinfo{author}{\bibfnamefont{W.-J.} \bibnamefont{Guo}},
  \bibinfo{author}{\bibfnamefont{B.-A.} \bibnamefont{Li}},
  \bibinfo{author}{\bibfnamefont{L.-W.} \bibnamefont{Chen}},
  \bibinfo{author}{\bibfnamefont{F.~J.} \bibnamefont{Fattoyev}},
  \bibnamefont{and} \bibinfo{author}{\bibfnamefont{W.~G.}
  \bibnamefont{Newton}}, \bibinfo{journal}{Physics Letters B}
  \textbf{\bibinfo{volume}{743}}, \bibinfo{pages}{408 } (\bibinfo{year}{2015}).

\bibitem[{\citenamefont{Li et~al.}(2018)\citenamefont{Li, Cai, Chen, and
  Xu}}]{Li18}
\bibinfo{author}{\bibfnamefont{B.-A.} \bibnamefont{Li}},
  \bibinfo{author}{\bibfnamefont{B.-J.} \bibnamefont{Cai}},
  \bibinfo{author}{\bibfnamefont{L.-W.} \bibnamefont{Chen}}, \bibnamefont{and}
  \bibinfo{author}{\bibfnamefont{J.}~\bibnamefont{Xu}}, \bibinfo{journal}{Prog.
  Part. Nucl. Phys.} \textbf{\bibinfo{volume}{99}}, \bibinfo{pages}{29 }
  (\bibinfo{year}{2018}).

\bibitem[{\citenamefont{Kong et~al.}(2017)\citenamefont{Kong, Xu, Chen, Li, and
  Ma}}]{Kong17}
\bibinfo{author}{\bibfnamefont{H.-Y.} \bibnamefont{Kong}},
  \bibinfo{author}{\bibfnamefont{J.}~\bibnamefont{Xu}},
  \bibinfo{author}{\bibfnamefont{L.-W.} \bibnamefont{Chen}},
  \bibinfo{author}{\bibfnamefont{B.-A.} \bibnamefont{Li}}, \bibnamefont{and}
  \bibinfo{author}{\bibfnamefont{Y.-G.} \bibnamefont{Ma}},
  \bibinfo{journal}{Phys. Rev. C} \textbf{\bibinfo{volume}{95}},
  \bibinfo{pages}{034324} (\bibinfo{year}{2017}).

\bibitem[{\citenamefont{Li and Han}(2013)}]{Li13}
\bibinfo{author}{\bibfnamefont{B.-A.} \bibnamefont{Li}} \bibnamefont{and}
  \bibinfo{author}{\bibfnamefont{X.}~\bibnamefont{Han}},
  \bibinfo{journal}{Phys. Lett. B} \textbf{\bibinfo{volume}{727}},
  \bibinfo{pages}{276} (\bibinfo{year}{2013}).

\bibitem[{\citenamefont{Holt et~al.}(2016)\citenamefont{Holt, Kaiser, and
  Miller}}]{Holt16}
\bibinfo{author}{\bibfnamefont{J.~W.} \bibnamefont{Holt}},
  \bibinfo{author}{\bibfnamefont{N.}~\bibnamefont{Kaiser}}, \bibnamefont{and}
  \bibinfo{author}{\bibfnamefont{G.~A.} \bibnamefont{Miller}},
  \bibinfo{journal}{Phys. Rev. C} \textbf{\bibinfo{volume}{93}},
  \bibinfo{pages}{064603} (\bibinfo{year}{2016}).

\bibitem[{\citenamefont{Baldo et~al.}(2017)\citenamefont{Baldo, Robledo,
  Schuck, and Vi\~nas}}]{Baldo17}
\bibinfo{author}{\bibfnamefont{M.}~\bibnamefont{Baldo}},
  \bibinfo{author}{\bibfnamefont{L.~M.} \bibnamefont{Robledo}},
  \bibinfo{author}{\bibfnamefont{P.}~\bibnamefont{Schuck}}, \bibnamefont{and}
  \bibinfo{author}{\bibfnamefont{X.}~\bibnamefont{Vi\~nas}},
  \bibinfo{journal}{Phys. Rev. C} \textbf{\bibinfo{volume}{95}},
  \bibinfo{pages}{014318} (\bibinfo{year}{2017}).

\bibitem[{\citenamefont{Mondal et~al.}(2017)\citenamefont{Mondal, Agrawal, De,
  Samaddar, Centelles, and Vi\~nas}}]{Mondal17}
\bibinfo{author}{\bibfnamefont{C.}~\bibnamefont{Mondal}},
  \bibinfo{author}{\bibfnamefont{B.~K.} \bibnamefont{Agrawal}},
  \bibinfo{author}{\bibfnamefont{J.~N.} \bibnamefont{De}},
  \bibinfo{author}{\bibfnamefont{S.~K.} \bibnamefont{Samaddar}},
  \bibinfo{author}{\bibfnamefont{M.}~\bibnamefont{Centelles}},
  \bibnamefont{and} \bibinfo{author}{\bibfnamefont{X.}~\bibnamefont{Vi\~nas}},
  \bibinfo{journal}{Phys. Rev. C} \textbf{\bibinfo{volume}{96}},
  \bibinfo{pages}{021302} (\bibinfo{year}{2017}).

\bibitem[{\citenamefont{Agrawal et~al.}(2017)\citenamefont{Agrawal, Samaddar,
  De, Mondal, and De}}]{Agrawal17}
\bibinfo{author}{\bibfnamefont{B.~K.} \bibnamefont{Agrawal}},
  \bibinfo{author}{\bibfnamefont{S.~K.} \bibnamefont{Samaddar}},
  \bibinfo{author}{\bibfnamefont{J.~N.} \bibnamefont{De}},
  \bibinfo{author}{\bibfnamefont{C.}~\bibnamefont{Mondal}}, \bibnamefont{and}
  \bibinfo{author}{\bibfnamefont{S.}~\bibnamefont{De}}, \bibinfo{journal}{Int.
  J. Mod. Phys.} \textbf{\bibinfo{volume}{E26}}, \bibinfo{pages}{1750022}
  (\bibinfo{year}{2017}).

\bibitem[{\citenamefont{Malik et~al.}(2018{\natexlab{b}})\citenamefont{Malik,
  Mondal, Agrawal, De, and Samaddar}}]{Malik18b}
\bibinfo{author}{\bibfnamefont{T.}~\bibnamefont{Malik}},
  \bibinfo{author}{\bibfnamefont{C.}~\bibnamefont{Mondal}},
  \bibinfo{author}{\bibfnamefont{B.~K.} \bibnamefont{Agrawal}},
  \bibinfo{author}{\bibfnamefont{J.~N.} \bibnamefont{De}}, \bibnamefont{and}
  \bibinfo{author}{\bibfnamefont{S.~K.} \bibnamefont{Samaddar}},
  \bibinfo{journal}{Phys. Rev. C} \textbf{\bibinfo{volume}{98}},
  \bibinfo{pages}{064316} (\bibinfo{year}{2018}{\natexlab{b}}).

\bibitem[{\citenamefont{Dobaczewski et~al.}(2014)\citenamefont{Dobaczewski,
  Nazarewicz, and Reinhard}}]{Dobaczewski14}
\bibinfo{author}{\bibfnamefont{J.}~\bibnamefont{Dobaczewski}},
  \bibinfo{author}{\bibfnamefont{W.}~\bibnamefont{Nazarewicz}},
  \bibnamefont{and} \bibinfo{author}{\bibfnamefont{P.-G.}
  \bibnamefont{Reinhard}}, \bibinfo{journal}{J. Phys. G: Nucl. Part. Phys.}
  \textbf{\bibinfo{volume}{41}}, \bibinfo{pages}{074001}
  (\bibinfo{year}{2014}).

\bibitem[{\citenamefont{Mondal et~al.}(2015)\citenamefont{Mondal, Agrawal, and
  De}}]{Mondal15}
\bibinfo{author}{\bibfnamefont{C.}~\bibnamefont{Mondal}},
  \bibinfo{author}{\bibfnamefont{B.~K.} \bibnamefont{Agrawal}},
  \bibnamefont{and} \bibinfo{author}{\bibfnamefont{J.~N.} \bibnamefont{De}},
  \bibinfo{journal}{Phys. Rev. C} \textbf{\bibinfo{volume}{92}},
  \bibinfo{pages}{024302} (\bibinfo{year}{2015}).

\bibitem[{\citenamefont{Chabanat et~al.}(1997)\citenamefont{Chabanat, Bonche,
  Haensel, Meyer, and Schaeffer}}]{Chabanat97}
\bibinfo{author}{\bibfnamefont{E.}~\bibnamefont{Chabanat}},
  \bibinfo{author}{\bibfnamefont{P.}~\bibnamefont{Bonche}},
  \bibinfo{author}{\bibfnamefont{P.}~\bibnamefont{Haensel}},
  \bibinfo{author}{\bibfnamefont{J.}~\bibnamefont{Meyer}}, \bibnamefont{and}
  \bibinfo{author}{\bibfnamefont{R.}~\bibnamefont{Schaeffer}},
  \bibinfo{journal}{Nucl. Phys.} \textbf{\bibinfo{volume}{A627}},
  \bibinfo{pages}{710} (\bibinfo{year}{1997}).

\bibitem[{\citenamefont{Wiringa et~al.}(1988)\citenamefont{Wiringa, Fiks, and
  Fabrocini}}]{Wiringa88}
\bibinfo{author}{\bibfnamefont{R.~B.} \bibnamefont{Wiringa}},
  \bibinfo{author}{\bibfnamefont{V.}~\bibnamefont{Fiks}}, \bibnamefont{and}
  \bibinfo{author}{\bibfnamefont{A.}~\bibnamefont{Fabrocini}},
  \bibinfo{journal}{Phys. Rev. C} \textbf{\bibinfo{volume}{38}},
  \bibinfo{pages}{1010} (\bibinfo{year}{1988}).

\bibitem[{\citenamefont{Wiringa}(1993)}]{Wiringa93}
\bibinfo{author}{\bibfnamefont{R.~B.} \bibnamefont{Wiringa}},
  \bibinfo{journal}{Rev. Mod. Phys.} \textbf{\bibinfo{volume}{65}},
  \bibinfo{pages}{231} (\bibinfo{year}{1993}).

\bibitem[{\citenamefont{Narikawa et~al.}(2018)\citenamefont{Narikawa, Uchikata,
  Kawaguchi, Kiuchi, Kyutoku, Shibata, and Tagoshi}}]{Narikawa18}
\bibinfo{author}{\bibfnamefont{T.}~\bibnamefont{Narikawa}},
  \bibinfo{author}{\bibfnamefont{N.}~\bibnamefont{Uchikata}},
  \bibinfo{author}{\bibfnamefont{K.}~\bibnamefont{Kawaguchi}},
  \bibinfo{author}{\bibfnamefont{K.}~\bibnamefont{Kiuchi}},
  \bibinfo{author}{\bibfnamefont{K.}~\bibnamefont{Kyutoku}},
  \bibinfo{author}{\bibfnamefont{M.}~\bibnamefont{Shibata}}, \bibnamefont{and}
  \bibinfo{author}{\bibfnamefont{H.}~\bibnamefont{Tagoshi}},
  \bibinfo{journal}{arXiv: 1812.06100}  (\bibinfo{year}{2018}).

\bibitem[{\citenamefont{Fujimoto et~al.}(2019)\citenamefont{Fujimoto,
  Fukushima, and Murase}}]{Fujimoto19}
\bibinfo{author}{\bibfnamefont{Y.}~\bibnamefont{Fujimoto}},
  \bibinfo{author}{\bibfnamefont{K.}~\bibnamefont{Fukushima}},
  \bibnamefont{and} \bibinfo{author}{\bibfnamefont{K.}~\bibnamefont{Murase}},
  \bibinfo{journal}{arXiv:1903.03400}  (\bibinfo{year}{2019}).

\bibitem[{\citenamefont{Fuchs}(2006)}]{Fuchs06}
\bibinfo{author}{\bibfnamefont{C.}~\bibnamefont{Fuchs}},
  \bibinfo{journal}{Progress in Particle and Nuclear Physics}
  \textbf{\bibinfo{volume}{56}}, \bibinfo{pages}{1 } (\bibinfo{year}{2006}),
  ISSN \bibinfo{issn}{0146-6410}.

\bibitem[{\citenamefont{Fantina et~al.}(2014)\citenamefont{Fantina, Chamel,
  Pearson, and Goriely}}]{Fantina14}
\bibinfo{author}{\bibfnamefont{A.~F.} \bibnamefont{Fantina}},
  \bibinfo{author}{\bibfnamefont{N.}~\bibnamefont{Chamel}},
  \bibinfo{author}{\bibfnamefont{J.~M.} \bibnamefont{Pearson}},
  \bibnamefont{and} \bibinfo{author}{\bibfnamefont{S.}~\bibnamefont{Goriely}},
  \bibinfo{journal}{EPJ Web of Conferences} \textbf{\bibinfo{volume}{66}},
  \bibinfo{pages}{07005} (\bibinfo{year}{2014}).

\bibitem[{\citenamefont{Danielewicz et~al.}(2002)\citenamefont{Danielewicz,
  Lynch, and Lacey}}]{Danielewicz02}
\bibinfo{author}{\bibfnamefont{P.}~\bibnamefont{Danielewicz}},
  \bibinfo{author}{\bibfnamefont{W.~G.} \bibnamefont{Lynch}}, \bibnamefont{and}
  \bibinfo{author}{\bibfnamefont{R.}~\bibnamefont{Lacey}},
  \bibinfo{journal}{Science} \textbf{\bibinfo{volume}{298}},
  \bibinfo{pages}{1592} (\bibinfo{year}{2002}).

\bibitem[{\citenamefont{Hebeler et~al.}(2013)\citenamefont{Hebeler, Lattimer,
  Pethick, and Schwenk}}]{Hebeler13}
\bibinfo{author}{\bibfnamefont{K.}~\bibnamefont{Hebeler}},
  \bibinfo{author}{\bibfnamefont{J.~M.} \bibnamefont{Lattimer}},
  \bibinfo{author}{\bibfnamefont{C.~J.} \bibnamefont{Pethick}},
  \bibnamefont{and} \bibinfo{author}{\bibfnamefont{A.}~\bibnamefont{Schwenk}},
  \bibinfo{journal}{Astrophys. J.} \textbf{\bibinfo{volume}{773}},
  \bibinfo{pages}{11} (\bibinfo{year}{2013}).

\bibitem[{\citenamefont{Colò et~al.}(2013)\citenamefont{Colò, Cao, Giai, and
  Capelli}}]{Colo13}
\bibinfo{author}{\bibfnamefont{G.}~\bibnamefont{Colò}},
  \bibinfo{author}{\bibfnamefont{L.}~\bibnamefont{Cao}},
  \bibinfo{author}{\bibfnamefont{N.~V.} \bibnamefont{Giai}}, \bibnamefont{and}
  \bibinfo{author}{\bibfnamefont{L.}~\bibnamefont{Capelli}},
  \bibinfo{journal}{Computer Physics Communications}
  \textbf{\bibinfo{volume}{184}}, \bibinfo{pages}{142 } (\bibinfo{year}{2013}),
  ISSN \bibinfo{issn}{0010-4655}.

\bibitem[{\citenamefont{Dutra et~al.}(2012)\citenamefont{Dutra,
  Louren\ifmmode~\mbox{\c{c}}\else \c{c}\fi{}o, S\'a~Martins, Delfino, Stone,
  and Stevenson}}]{Dutra12}
\bibinfo{author}{\bibfnamefont{M.}~\bibnamefont{Dutra}},
  \bibinfo{author}{\bibfnamefont{O.}~\bibnamefont{Louren\ifmmode~\mbox{\c{c}}\else
  \c{c}\fi{}o}}, \bibinfo{author}{\bibfnamefont{J.~S.}
  \bibnamefont{S\'a~Martins}},
  \bibinfo{author}{\bibfnamefont{A.}~\bibnamefont{Delfino}},
  \bibinfo{author}{\bibfnamefont{J.~R.} \bibnamefont{Stone}}, \bibnamefont{and}
  \bibinfo{author}{\bibfnamefont{P.~D.} \bibnamefont{Stevenson}},
  \bibinfo{journal}{Phys. Rev. C} \textbf{\bibinfo{volume}{85}},
  \bibinfo{pages}{035201} (\bibinfo{year}{2012}).

\end{thebibliography}

\end{document}